\documentclass[twocolumn,prl,aps,superbib,tightenlines,floatfix,superscriptaddress,showpacs]{revtex4}
\usepackage{amsmath}
\usepackage{graphicx}
\usepackage{amssymb}
\usepackage{color,soul}
\usepackage[T1]{fontenc}
\usepackage{ae,aecompl}
\usepackage{color,soul}

\begin{document}

\title{Reconfigurable Transmission Lines with Memcapacitive Materials}%

\author{Y. V. Pershin}
\email{pershin@physics.sc.edu} \affiliation{Department of Physics and Astronomy and
Smart State Center for Experimental Nanoscale Physics, University of South Carolina, Columbia, South Carolina 29208, USA}
\author{V. A. Slipko}
\affiliation{Department of Physics and Technology, V. N. Karazin Kharkov National University, Kharkov 61022, Ukraine}
\author{M. Di Ventra}
\email{diventra@physics.ucsd.edu} \affiliation{Department of Physics, University of California San Diego, La Jolla, California 92093, USA}

\begin{abstract}
We study transmission lines made of memory capacitive (memcapacitive) materials. The transmission properties of these lines can be adjusted {\it on demand} using an appropriate sequence of pulses. In particular, we demonstrate a pulse combination that creates a periodic modulation of dielectric properties along the line. Such a structure resembles a distributed Bragg reflector having important optical applications. We present simulation results demonstrating all major steps of such a reconfigurable device operation including reset, programming and transmission of small amplitude signals. The proposed {\it reconfigurable transmission lines} employ only passive memory materials and can be realized using
available memcapacitive devices.
\end{abstract}

\maketitle

Transmission lines are useful and ubiquitous structures developed for a wide range of applications ranging from the transmission of radio frequency and microwave signals \cite{book:1381215} to coupling of superconducting qubits \cite{Wallraff04a,Majer07a}, to name a few.  In particular, coaxial cables are  transmission lines for radio frequency signals. Their design consists in a shielded central core separated from a shield by a dielectric material \cite{Goleniewski06a}. More complex transmission line designs can be realized with metamaterials \cite{book:450803,book:1381215}.

In general, transmission lines are built to support specific transmission characteristics (such as the frequency range, etc.)
 that are fixed once and for all by their design. In other words, in order to change, e.g., their frequency range one needs to change the materials (or structure) of the line itself. It would be very beneficial if instead we could {\it reconfigure} the transmission properties {\it on demand} by a simple application of appropriate input signals. This way, a single transmission line could perform different tasks functioning, for example, as a delay line, band-stop filter, etc.

\begin{figure*}[htp]
 \includegraphics[angle=0,width=\textwidth]{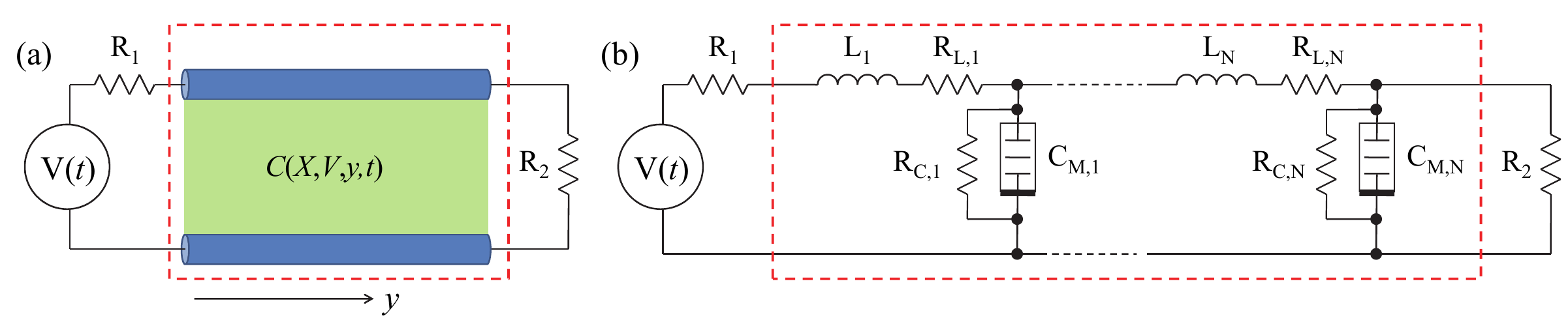}
\caption{(Color online) (a) Memcapacitive transmission line (the area enclosed by a dashed line) connected to the voltage source $V(t)$ from the left and load resistor $R_2$ from the right. Here, R$_1$ represents a combination of the matching resistance and internal resistance of $V(t)$. (b) The same circuit as in (a) with memcapacitive transmission line represented by its circuit model, where L$_i$ are inductors, C$_{\textnormal{M},i}$ are memcapacitive systems, R$_{\textnormal{L},i}$ and R$_{\textnormal{C},i}$ are resistors used to simulate loss, $i=1,..,N$. \label{fig1}}
\end{figure*}

In this paper, we propose precisely this concept by introducing {\it reconfigurable transmission lines} based on memcapacitive materials \cite{diventra09a,diventra11a}. Such lines could be realized in practice by, e. g., partially (or entirely) replacing the usual dielectric insulator in a coaxial cable (or other transmission line realization) with a memcapacitive material, namely, a material whose relative permittivity depends on the history of signals applied (see Fig. \ref{fig1}(a) for a conceptual image)~\cite{diventra11a}. Using an appropriate combination of pulses, one can pre-program the properties of memcapacitive materials on demand, and thus select the function that the transmission line performs. Moreover, our idea could be realized as an electronic circuit involving memcapacitive devices, resistors and inductors. In fact, the simulations presented below are based on a circuit model of the transmission line \cite{book:1381215}. Therefore, our results are applicable to both the line itself and its electronic circuit realization.

Let us consider the transmission line positioned along the $y$-axis as shown in Fig. \ref{fig1}(a). We assume that the electromagnetic field of the line satisfies the quasi-stationarity conditions with  respect to the transverse dimensions of the line (in particular,
the wavelength of the electromagnetic field in the line should be much longer than the line's transverse dimensions).
Using Maxwell's equations,
we can get the following well-known equations of the transmission line
\begin{eqnarray}
\frac{\partial \Phi}{\partial t}+RI+\frac{\partial V}{\partial y}=0,
\label{eq:flux}\\
 \frac{\partial q}{\partial t}+GV+\frac{\partial I}{\partial y}=0,
\label{eq:charge}
\end{eqnarray}
where $\Phi=LI$ is the magnetic field flux, $L$ is the inductance,
$R$ is the longitudinal resistivity, $C=q/V$ is the capacitance, and $G$ the transverse conductivity
(all these quantities are defined per unit length of the line in the $y$ direction). Moreover,
$I$ is the current in the top wire, $V$ is the voltage
across the line, and $q$ is the 1D charge density in the top wire.
In Eq. (\ref{eq:charge}) it is assumed that the leakage current is directly proportional
to the voltage $V$.

Next, let us consider the memory feature of our transmission line. Generally speaking, any of the line's parameters
($R$, $G$, $C$ or $L$) could depend on some (dynamic) internal state variables (fields) and thus describe the memory response \cite{diventra09a}.
In this paper, however, we focus solely on the memory in the line's capacitance, $C(X,V,y,t)$, that can be easily
implemented experimentally. In fact, there are several possible realizations of memcapacitive systems described in the literature \cite{pershin11a,liu06a,martinez09a,driscoll09a,Lai09a,pershin11c,Flak14a}. However, we do not refer to anyone specific system here, so as to keep the discussion general.
Importantly, the line's capacitance depends on internal state variables (fields) $X$ with their own equations of motion
\begin{eqnarray}
\frac{\partial X(y,t)}{\partial t}=F(X,V,y,t),
\label{eq:X}
\end{eqnarray}
 where $X$ is a vector-function, and in general $F$ is a functional describing the memory capacitance mechanism of the specific material.

Fig. \ref{fig1}(b) presents the circuit model of memcapacitive transmission line. This model can be obtained using a finite-difference discretization
of Eqs. (\ref{eq:flux}) and (\ref{eq:charge}) with respect to the $y$-coordinate:
\begin{eqnarray}
\frac{\partial \Phi_i}{\partial t}+R_{L,i}I_{i}+V_{i}-V_{i-1}=0,
\label{eq:flux_i}\\
 \frac{\partial q_{i}}{\partial t}+G_{C,i}V_{i}+I_{i+1}-I_{i}=0.
\label{eq:charge_i}
\end{eqnarray}
Here,
$\Phi_i=L_i I_i$, $L_i=\Delta y L$, $R_{L,i}=\Delta y R$,
 $q_i=C_{M,i}(x_i,V_i,t)V_i$, $G_{C,i}=\Delta y G\equiv R_{C,i}^{-1}$, and $\Delta y$ is the discretization step. The memcapacitance $C_{M,i}(x_i,V_i,t)=\Delta y C(X,V,y_i,t)$ \cite{diventra09a} is essentially a generalized capacitance with memory. A broad class of memcapacitive systems and devices (so-called $n$th-order voltage-controlled memcapacitive systems \cite{diventra09a}) is defined by the following general equations \cite{diventra09a}
\begin{eqnarray}
q(t)&=&C(x,V_C,t)V_C(t) \label{eq:def1} \\
\dot{x}&=&f(x,V_C,t), \label{eq:def2}
\end{eqnarray}
where $q(t)$ is the charge on the device at time $t$, $V_C(t)$ is the voltage across the memcapacitive system, $C(x,V_C,t)$ is the memcapacitance, $x$ is an $n$-component vector of internal state variables, and $f(x,V_C,t)$ is an $n$-dimensional vector function.

\begin{figure*}[t]
\begin{center}
 \includegraphics[angle=0,width=3.3cm]{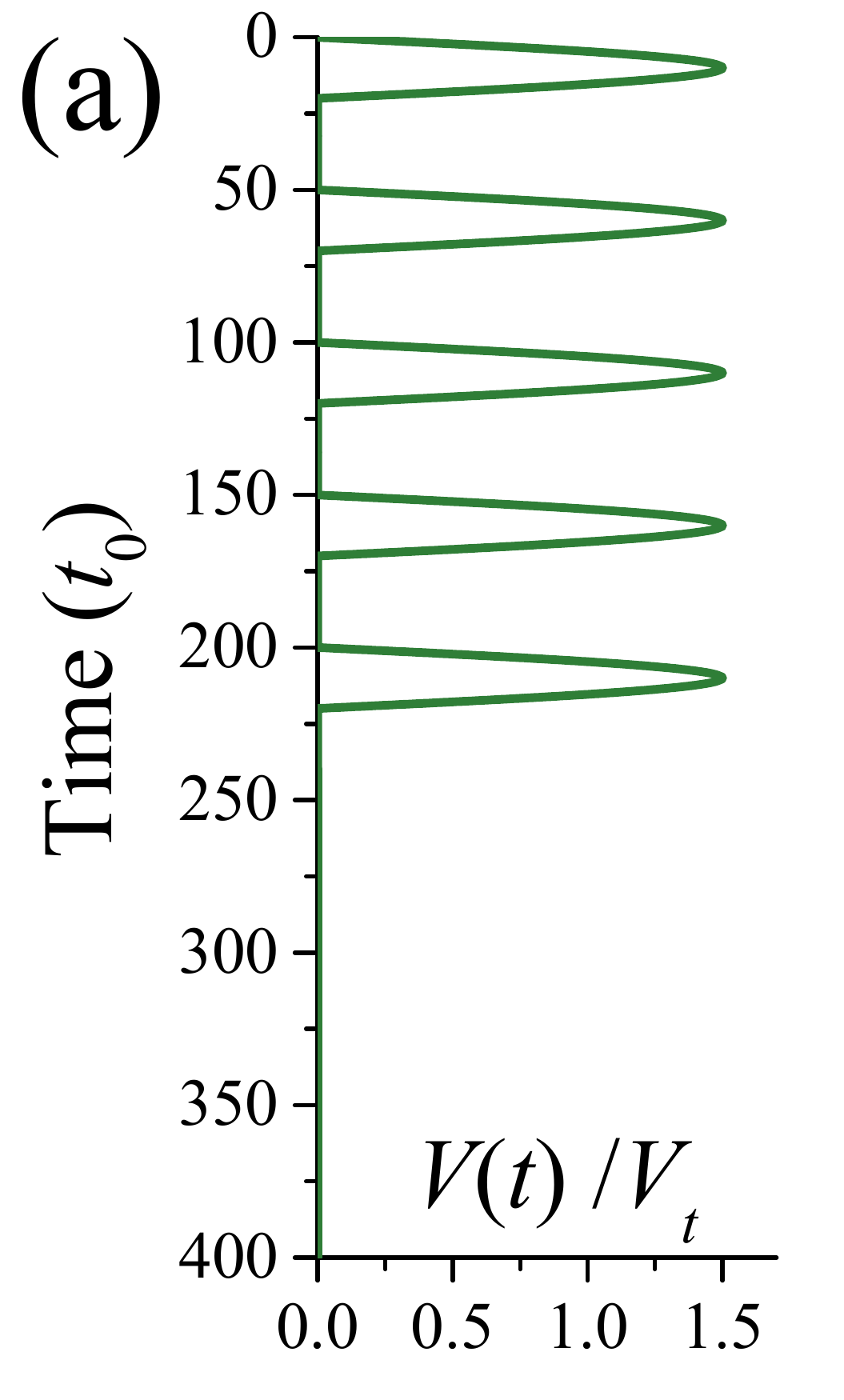}
 \includegraphics[angle=0,width=7.5cm]{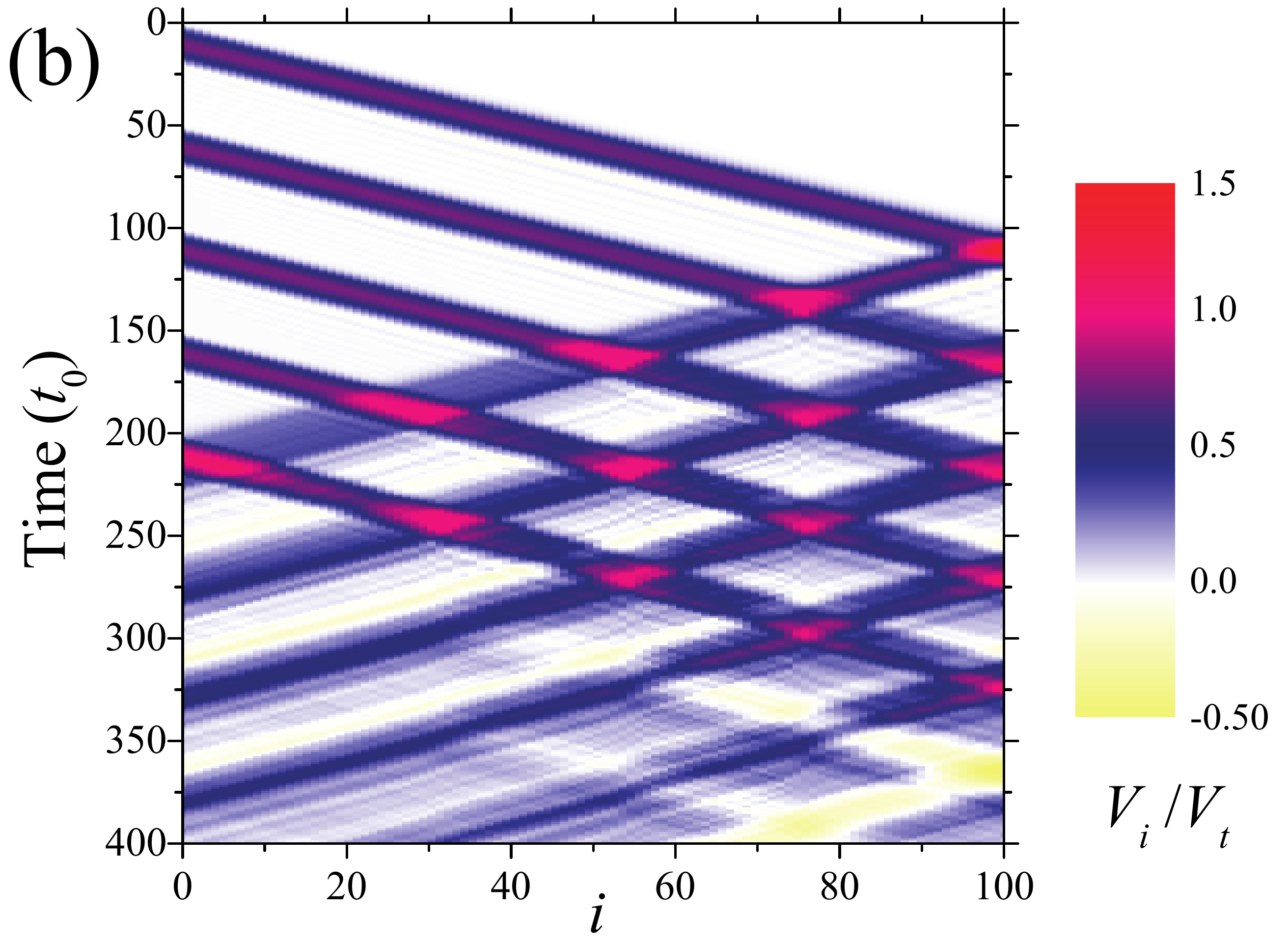}
 \includegraphics[angle=0,width=6.5cm]{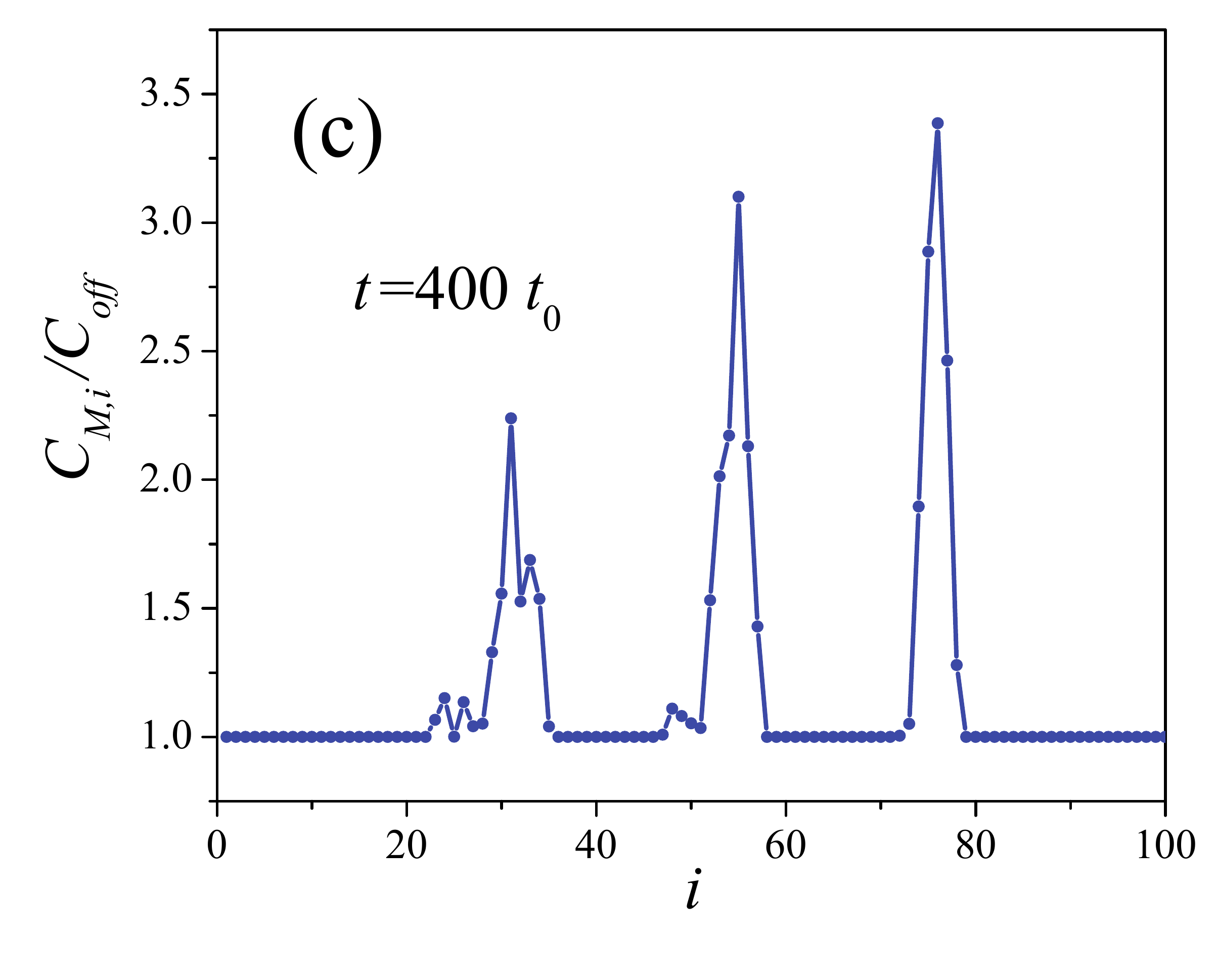}
\caption{(Color online) (a) Shape of the voltage pulses produced by the voltage source $V(t)$. Five pulses are emitted with $50t_0$ time delay, where $t_0$ is the unit of time. Each pulse is a half-period of sine wave with a width of $20t_0$ and amplitude $1.5V_t$. (b) Propagation of five input pulses through the transmission line with $R_1=Z_0$ and $R_2\gg Z_0$ (almost perfect reflection at the line's end and no reflection at the beginning for the reflected signals). The line consists of 100 units, the units with $i<10$ and $i>90$ employ regular capacitors with capacitance $C_{off}$. This simulation was performed using the following set of parameters: $C_{on}/C_{off}=5$, $R_1C_{off}/t_0=1$, $R_2C_{off}/t_0=10^4$, $R_{L,i}C_{off}/t_0=10^{-3}$, $R_{C,i}C_{off}/t_0=10^6$, $V_{tr}\beta t_0=50$. Note that the input signal magnitude drops by one half on $R_1$. (c) Final distribution of memcapacitances along the line. Noteworthy are the peaks that correspond to the areas of pulse overlaps in (b). \label{fig2}}
\end{center}
\end{figure*}

For our purposes and since they are the most common one, it is convenient to employ memcapacitive materials with threshold-type switching characteristics. A model of bipolar voltage-controlled memcapacitive device with threshold \cite{diventra09a} could be used to represent the properties of desirable materials in the circuit model of the transmission line. Such threshold-type models describe devices whose capacitance changes only if the voltage magnitude across the device, $V_C$, exceeds the threshold voltage, $V_t$, namely, $|V_C|>V_t$. Moreover, the direction of capacitance change is defined by the voltage polarity. Specifically, in our simulations we use the following model of memcapacitive system with threshold:
\begin{eqnarray}
C(x)&=&C_{off}+x(C_{on}-C_{off}) \label{eq:1} \\
\frac{\textnormal{d}x}{\textnormal{d}t}&=&\left\{ \begin{array}{cc} \textnormal{sign}(V_C)\beta\left( |V_C|-V_t\right) & \; \textnormal{if} \; |V_C|\geq V_t \\ 0 & \textnormal{otherwise} \end{array} , \right. \label{eq:2} \\ \nonumber
\end{eqnarray}
where the memcapacitance  $C(x)$ changes continuously between $C_{off}$ and $C_{on}$, $C_{on}>C_{off}$, $x$ is the internal state variable taking values in the range between 0 and 1, $\beta$ is the constant defining the rate of change of $x$, and $sign\{\cdot\}$ is the sign of the argument. According to the above equations, $C(x)$ is increased at positive applied voltages and decreased at negative ones provided that the voltage magnitude exceeds $V_t$.

The signal propagation in the ideal lossless transmission lines composed of linear inductances $L$ and capacitances $C_{off}$ is described by the wave equation with the velocity given by $v=1/\sqrt{LC_{off}}$. In such lines, the excitation propagates unaltered along the line and its reflection from the load resistance is defined by the relation between the line impedance $Z_0=\sqrt{L/C_{off}}$ and the load resistance $R_2$.
In particular, the excitation is completely reflected at $R_2\ll Z_0$ and $R_2\gg Z_0$ with and without sign reversal, respectively. The reflection is minimized at $R_2=Z_0$.

\begin{figure}[t]
\begin{center}
 \includegraphics[angle=0,width=7cm]{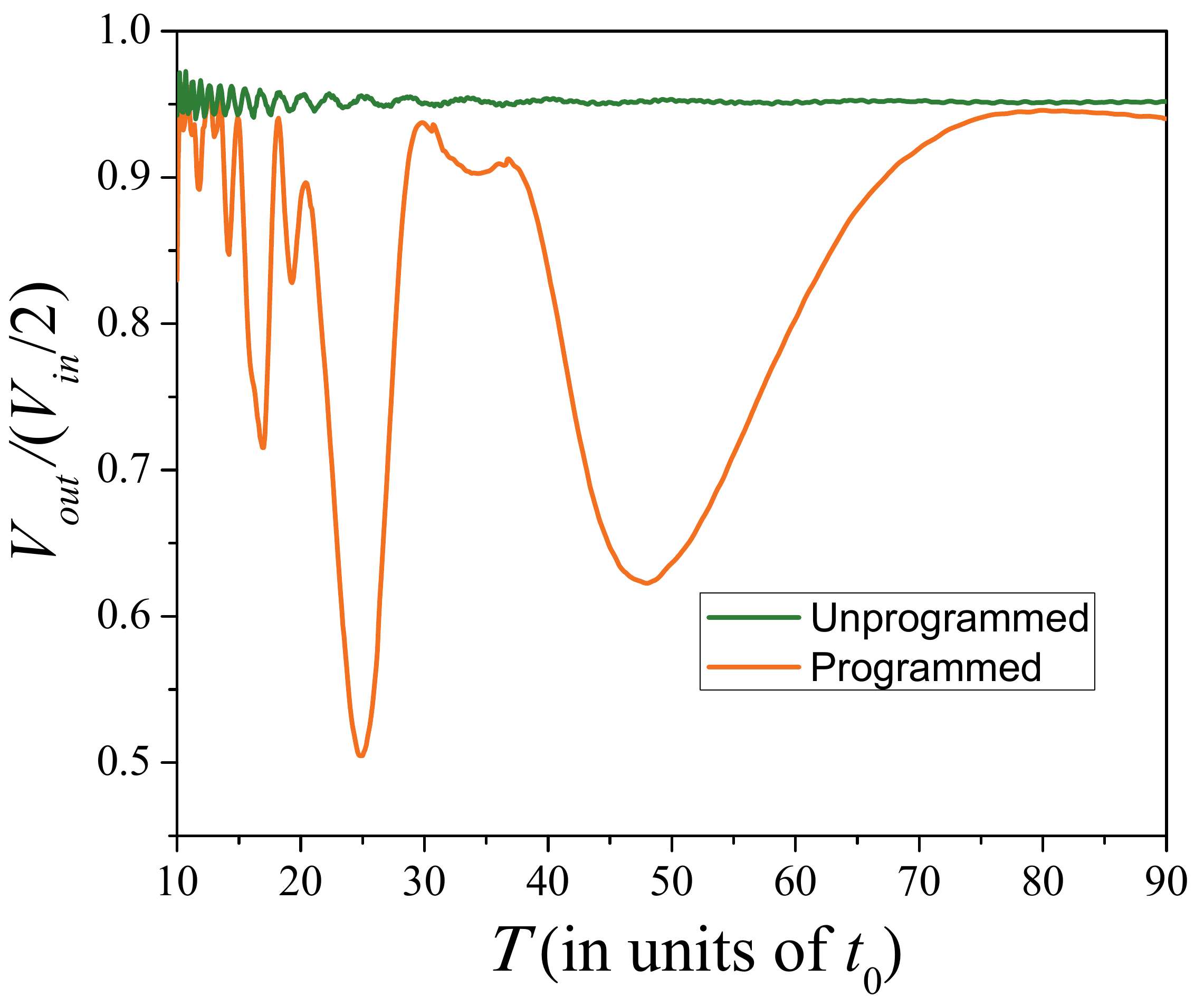}
\caption{Reduction in the transmitted $ac$ signal amplitude (the $ac$ voltage amplitude across $R_2$) as a function of the $ac$ period $T$ for the initial (flat) and final (shown in Fig. \ref{fig2}(b)) configurations of memcapacitances. \label{fig3}}
\end{center}
\end{figure}

In the following, we consider the three major steps of operation of a general reconfigurable transmission line.  Figs. \ref{fig2} and \ref{fig3}  exemplify
these steps based on the discrete model of the transmission line shown in Fig. \ref{fig1}(b). In order to obtain Figs. \ref{fig2} and \ref{fig3},  we set up and solved a system of equations based on Kirchoff's rules (coinciding with Eqs. (\ref{eq:flux_i})-(\ref{eq:charge_i}) for the bulk of transmission line)
supplemented by Eqs.  (\ref{eq:1})-(\ref{eq:2}) for the evolution of memcapacitive systems. As the initial condition, we assume that there is no energy stored in the circuit. The list of specific circuit parameters is given in Fig. \ref{fig2} caption. 

{\it Reset.}-- The reset of the memcapacitive transmission line is a straightforward task. The purpose of this stage is to set all memcapacitive systems into the same state. For the sake of definiteness, let us consider the low capacitance state $C_{off}$ as the desirable state after the reset. In order to reach this state, one can apply (sufficiently slow) a constant negative voltage to the line such that the voltage across all memcapacitive systems exceeds their threshold. Similarly, by applying a constant positive voltage to the line one can set all the capacitances into $C_{on}$.

{\it Programming.}--In order to program the desired distribution of memcapacitances along the line, we suggest the application of time-delayed pulses under the condition of pulse reflection from the far end of the line. Such a condition (without sign reversal) is achieved by selecting $R_2\gg Z_0$. Our main idea is to use the pulse overlap to locally create strong electric fields thus limiting the capacitance switching to only these strong field regions. In this approach, the pulse delay is used as a control parameter defining the switching position. It should be emphasized that the amplitudes of individual pulses need to be kept below the threshold voltage.

A simulation of the programming step is presented in Fig. \ref{fig2}. In this simulation it is assumed that the first and last 10 units of the transmission line (the total number of units $N=100$) are made of regular capacitors of capacitance $C_{off}$. The voltage source $V(t)$ generates five right-moving pulses (Fig. \ref{fig2}(a)) propagating along the line (the pulse separation is $50t_0$, where $t_0=\sqrt{LC_{off}}$). The overlap of the right-moving and reflected left-moving pulses develops a strong electric field switching (increasing) the memcapacitance in the areas of overlap (Fig. \ref{fig2}(b)). According to the selected pulse delay and line configuration, there are three regions of the increased capacitance as seen in Fig. \ref{fig2}(c). Alternatively, the desired memcapacitance distribution could be achieved by using two time-delayed voltage sources connected to the opposite ends of the line while maintaining non-reflective boundary conditions at both ends.

{\it Transmission of small amplitude signals.}--Most of the time, reconfigurable transmission lines will be used to transfer small amplitude signals. During this operation step the voltage magnitude at any point of the line should always be smaller than the threshold voltage of the memcapacitive systems to keep their states unchanged. The distribution of memcapacitance along the line created during the programming step thus defines the line functionality. In our example of the programming step presented in Fig. \ref{fig2}, the final distribution of memcapacitance exhibits some periodic variation similar to those in distributed Bragg reflectors \cite{book:2444,B903229K}. Each variation of memcapacitance causes a partial reflection of an incoming {\it ac} wave with a constructive interference for waves having a wavelength close to the periodicity of memcapacitance variation.

Fig. \ref{fig3} presents simulations of small amplitude {\it ac} signal propagation through the line unprogrammed and programmed as in Fig.~\ref{fig2} as a function of the signal wavelength. This plot clearly shows two major dips for wavelengths of about 25 and 50 (in units of $t_0$) corresponding to the periodicity of the memcapacitance variation as shown in Fig. \ref{fig2}(c). The corresponding wavelengths correspond to the usual stop-bands of distributed Bragg reflectors \cite{book:2444,B903229K}.

In conclusion, we have proposed reconfigurable transmission lines based on memcapacitive systems. Such lines are capable of performing signal processing functions on demand depending on the memcapacitance distribution along the line which can be pre-programmed with appropriate pulses.  Few years ago, an analog memcapacitive device based on field-configurable ion-doped polymers was experimentally demonstrated \cite{Lai09a}. Such device offers an example of the structures that can be used to build discrete or continuum versions of the reconfigurable memcapacitive transmission lines proposed here. Since the capacitance of ion-doped polymers remains practically unchanged for extended periods of time under continual reading conditions by low amplitude signals \cite{Lai09a}, it is expected that the general characteristics of low-signal response of reconfigurable lines (e.g., their bandwidth) are similar to those for traditional lines employing the same but fixed properties materials. We expect such reconfigurable transmission lines to find applications in reconfigurable networks, as programmable delay lines, band-stop filters, etc.

This work has been partially supported by NSF grant ECCS-1202383, Smart State Center for Experimental Nanoscale Physics, and the Center for Magnetic Recording Research at UCSD.

\bibliography{memcapacitor}

\end{document}